\documentclass[conference]{IEEEtran}
    \usepackage{cite}
    \usepackage[english]{babel}
    \usepackage[T1]{fontenc}
    \usepackage[utf8]{inputenc} 
    \usepackage{amsmath,amssymb,amsfonts}
    \usepackage{algorithmic}
    \usepackage{booktabs}
    \usepackage{tabularx}
    \usepackage{graphicx}
    \usepackage{textcomp}
    \usepackage{color}
    \usepackage{glossaries}
    \usepackage{enumitem}
    \usepackage{subcaption}
    \usepackage{placeins}
    \usepackage{flushend}
    \usepackage{soul}
    \usepackage{xcolor}
    \usepackage{mdframed}
    \usepackage{lipsum}
    \usepackage{pgf}
    \usepackage{pgfplots}
    \usepgfplotslibrary{external} 
    \usepackage{soul}
    \usepackage[printwatermark]{xwatermark}
    \usepackage{siunitx}
    \DeclareSIUnit{\belmillii}{Bi}
    \DeclareSIUnit{\belmilliwatt}{Bm}
    \DeclareSIUnit{\dBm}{\deci\belmilliwatt}
    \DeclareSIUnit{\dBi}{\deci\belmillii}
    \usepackage{url}
    \usepackage[numbers]{natbib}

\makeglossaries
\newacronym{iot}{IoT}{Internet of Things}
\newacronym{lorawan}{LoRaWAN}{Long Range Wide Area Network}
\newacronym{lora}{LoRa}{Long Range}
\newacronym{p2p}{P2P}{Point-to-Point}
\newacronym{rss}{RSS}{Received Signal Strength}
\newacronym{css}{CSS}{Chirp Spread Spectrum}
\newacronym{adr}{ADR}{Adaptive Data Rate}
\newacronym{erp}{ERP}{Effective radiated power}
\newacronym{ide}{IDE}{Integrated Development Environment}
\newacronym{lbt}{LBT}{Listen Before Talk}
\newacronym{afa}{AFA}{Adaptive Frequency Agility}
\newacronym{snr}{SNR}{signal-to-noise ratio}
\newacronym{sf}{SF}{Spreading Factor}
\newacronym{lpwan}{LPWAN}{Low Power Wide Area Network}
\newacronym{pdr}{PDR}{packet delivery ratio}
\newacronym{pl}{PL}{Path Loss}
\newacronym{los}{LoS}{Line-of-Sight}

\def\BibTeX{{\rm B\kern-.05em{\sc i\kern-.025em b}\kern-.08em
    T\kern-.1667em\lower.7ex\hbox{E}\kern-.125emX}}

\IEEEoverridecommandlockouts
\newcommand*{\nolink}[1]{%
  {\protect\NoHyper#1\protect\endNoHyper}%
} 
\PassOptionsToPackage{bookmarks={false}, hyperfootnotes={false}}{hyperref}
\begin{document}

\title{LoRa Physical Layer Evaluation for Point-to-Point Links and Coverage Measurements in Diverse Environments\\
\thanks{This research is supported by the EU Horizon 2020 program IoF2020, grant nr. 731884, IoTRAILER Use Case.}
\thanks{This paper is accepted and published in the Conference Proceedings of 2019 European Conference on Networks and Communications (EuCNC).}
}

\author{\IEEEauthorblockN{Gilles Callebaut, Guus Leenders, Chesney Buyle, Stijn Crul, Liesbet Van der Perre}
\IEEEauthorblockA{\textit{KU Leuven, DRAMCO, Department of Electrical Engineering (ESAT)} \\
\textit{Ghent Technology Campus,}\\
B-9000 Ghent, Belgium \\
name.surname@kuleuven.be}
}

\maketitle
\begin{abstract}
This paper focuses on \acrlong{p2p} LoRa connections. We report on coverage measurements performed in three distinct environments, i.e., coastal, forest and urban. Our field experiments demonstrate coverage up to \SI{1}{\kilo\meter} with antennas at only \SI{1.5}{\meter} height in an urban scenario. 
In free \gls{los} this coverage is extended to \SI{4}{\kilo\meter}. 
Based on these results, we are developing a path loss model in future work. The developed hardware and software including measurements are open-source.
\end{abstract}


\section{Introduction and State of the Art}
\gls{lora} technology in an interesting candidate to connect \gls{iot} applications requiring medium to long range connectivity. The operation in an unlicensed band opens the opportunity to establish ad-hoc networks and \gls{p2p} connections, avoiding significant network infrastructure investments or subscription costs.
We here evaluate the behaviour of the \gls{lora} physical layer with low-height terminals enabling such deployments.
Measurements were performed in significantly varied environments: urban, coastal and densely forested.  
Several studies have considered \gls{lorawan} network coverage  and network performance \nolink{\cite*{Petajajarvi2015, PetajajarviJuha2016EoLL, 10.1007/978-3-319-48799-1_21}}. 
We here focus on \gls{lora} \gls{p2p} in outdoor environments, leveraging  on a LoRa concentrator mounted at a height of approximately \SI{1.5}{\meter}, while a LoRaWAN gateway is typically placed high above ground.
 
\begin{figure}[]
    \centering
    \includegraphics[width=0.8\linewidth]{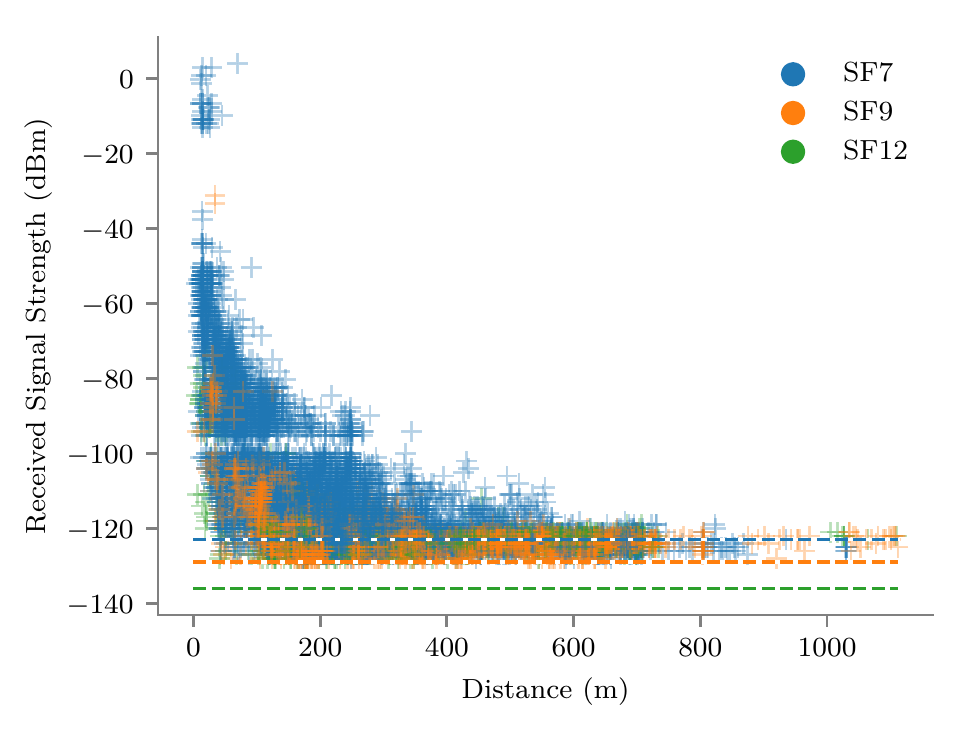}    
    \caption{\gls{rss} with respect to the Rx-Tx separation for the urban scenario. The \gls{rss} sensitivity of the \gls{lora} demodulator is presented by a dashed line for each \gls{sf}.}%
    \label{fig:rss}
\end{figure}

\section{LoRa P2P System and Set-Up}\label{sec:lora-vs-lorawan}
In respect to the OSI model, \gls{lora} is the physical layer and \gls{lorawan} extends to the datalink and the network layer. Long range connections are achieved by exploiting a coding gain originating from the used spread spectrum technique of the \gls{lora} modulation. The coding gain is proportional to the \gls{sf}. By increasing the \gls{sf}, the range can be extended due to the lowered \gls{snr} and \gls{rss} demodulation floor. Reducing the spreading factor results in a lower time on air. Thereby, allowing to trade-off range and power consumption. 
Employing just the \gls{lora} PHY layer to create \gls{p2p} networks offers some distinct advantages over utilizing \gls{lorawan}. 
Expensive gateways and back-ends can be avoided, specific frequency bands can be assigned to specific devices thus mitigating interference. In its simplest form, the devices can be configured to send-and-forget allowing for straightforward setup if the application is already developed. This yields power improvements and lower device expenditure. Additionally, devices do not need to be registered to a network prior to sending packets, contrary to a deployment in a \gls{lorawan}.
As the \gls{lorawan} specification~\nolink{\cite{lora:1.1}} is not implemented, \gls{lora} \gls{p2p} devices can exploit channel 6 otherwise used by \gls{lorawan} gateways for downlink communication. The high duty cycle and a higher allowed transmit power is sometimes favorable over the other available channels. Furthermore, \gls{lorawan} gateways make use of IQ reversal to mitigate interference between uplink and downlink communication. As a consequence, deploying \gls{p2p} devices operating in channel 6 should be subjected to less interference in the presence of \gls{lorawan} networks. To be complete, the permitted airtime may exceed the duty cycle limit when applying \gls{lbt} and \gls{afa}, but this will yield a higher power consumption. 

We examined the point-to-point \gls{lora} links based on our in-house developed platform hosting the popular SX1276~\nolink{\cite{semtechsx127x}} transceiver. 
The measurements were conducted with three stationary transmitters locked to each transmit at a specific \gls{sf}, respectively SF7, SF9 and SF12. They are positioned horizontally at a distance of $2\lambda$ relative to each other to mitigate antenna coupling, at approximately \SI{1.5}{m} above ground. The portable receivers log both the received RF parameters and the corresponding position.

\begin{figure}[h]
    \centering
    \begin{subfigure}[b]{0.42\linewidth}
        \centering
        \includegraphics[width=0.9\linewidth]{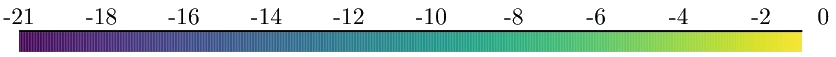}
    \end{subfigure}
    \begin{subfigure}[b]{0.42\linewidth}
        \centering
        \includegraphics[width=0.9\linewidth]{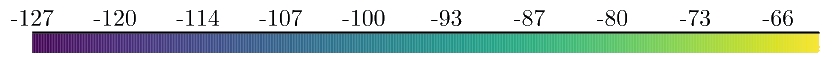}
    \end{subfigure} 
    ~%
    \begin{subfigure}[b]{0.42\linewidth}
        \centering
        \includegraphics[width=0.9\linewidth]{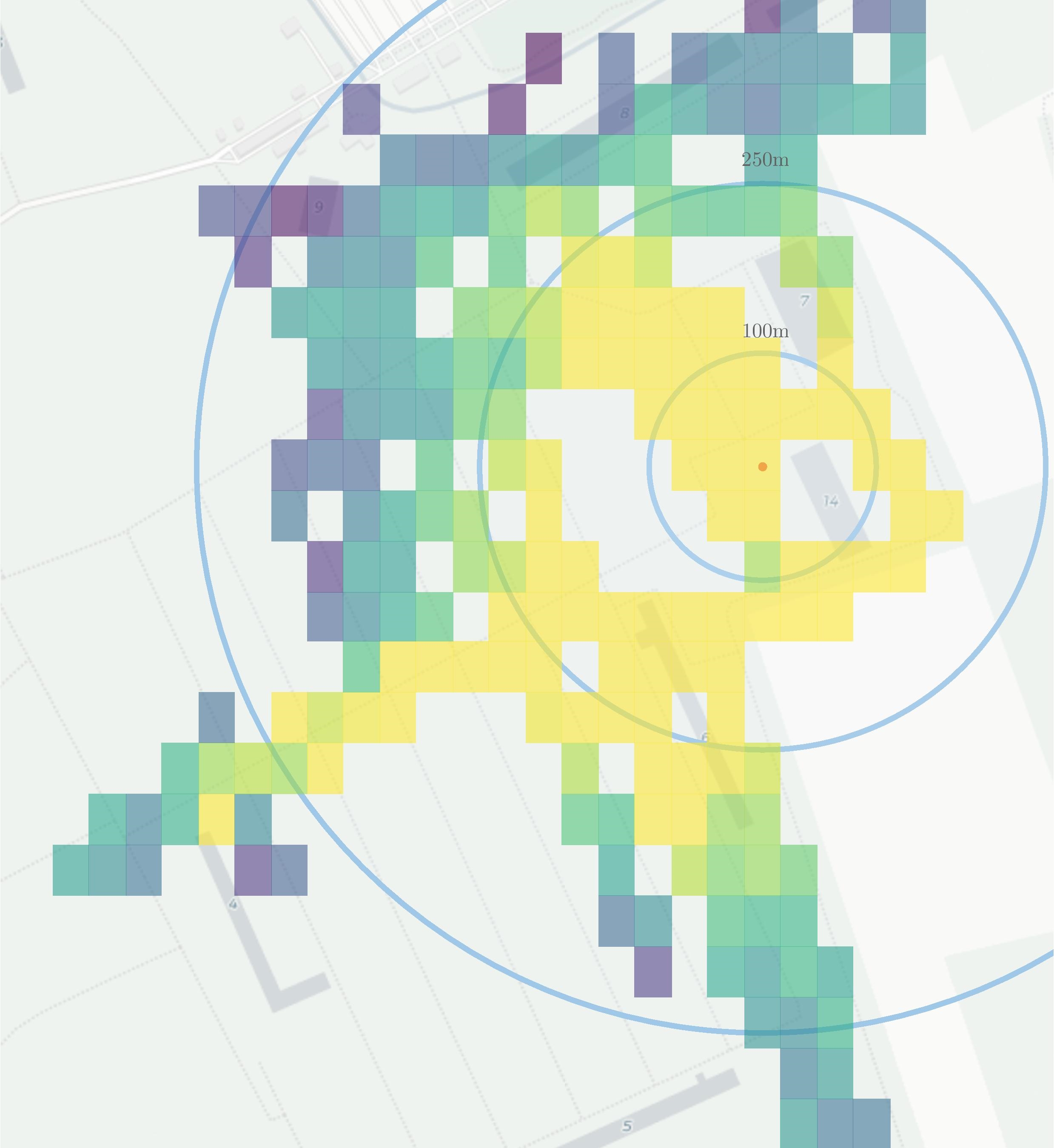}
        \caption{Forest SNR}%
        \label{fig:heatmap-snr-forest}
    \end{subfigure}
    \begin{subfigure}[b]{0.42\linewidth}
        \centering
        \includegraphics[width=0.9\linewidth]{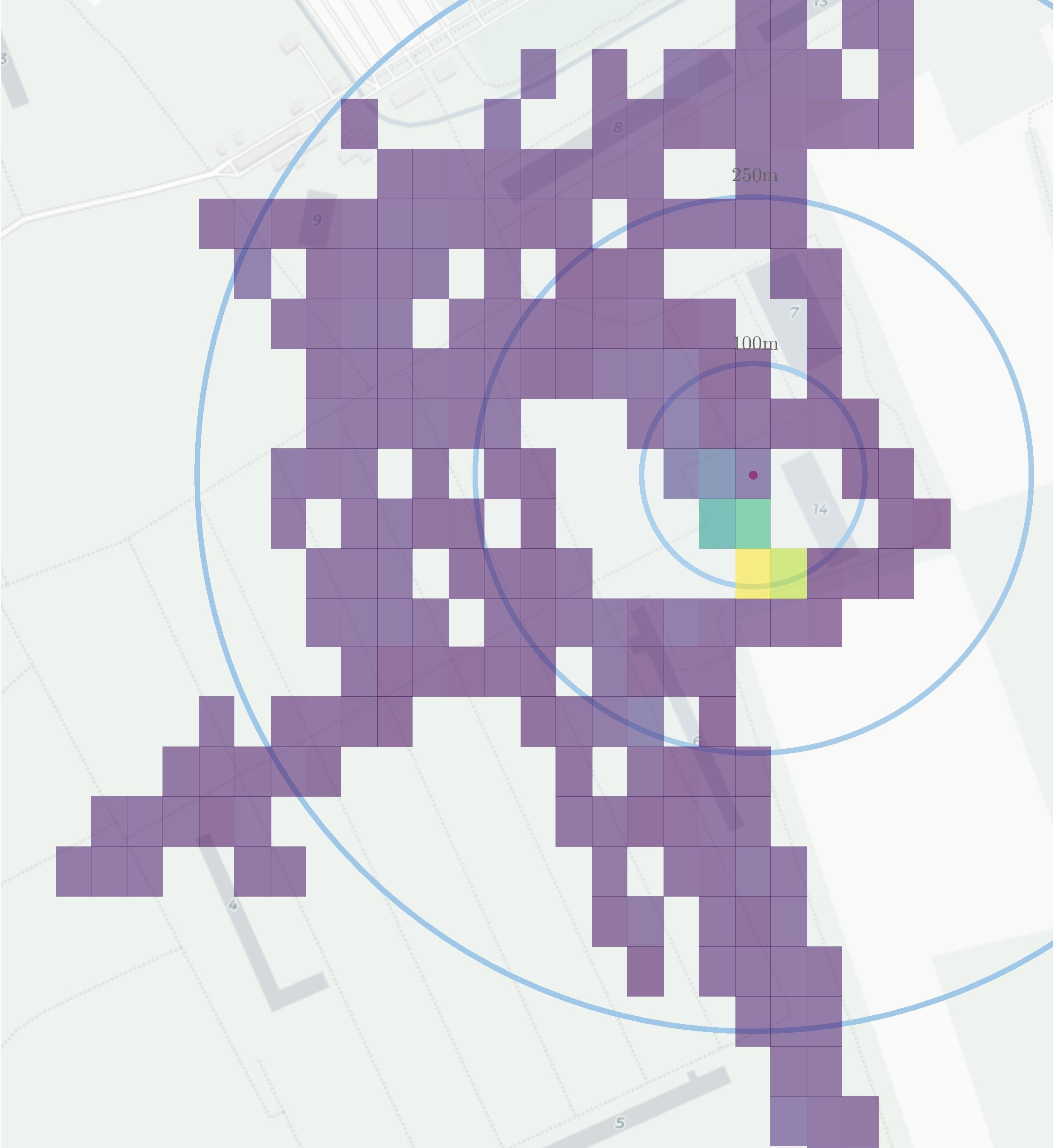}
        \caption{Forest RSS}%
        \label{fig:heatmap-rss-forest}
    \end{subfigure} 
    ~%
    \begin{subfigure}[b]{0.42\linewidth}
        \centering
        \includegraphics[width=0.9\linewidth]{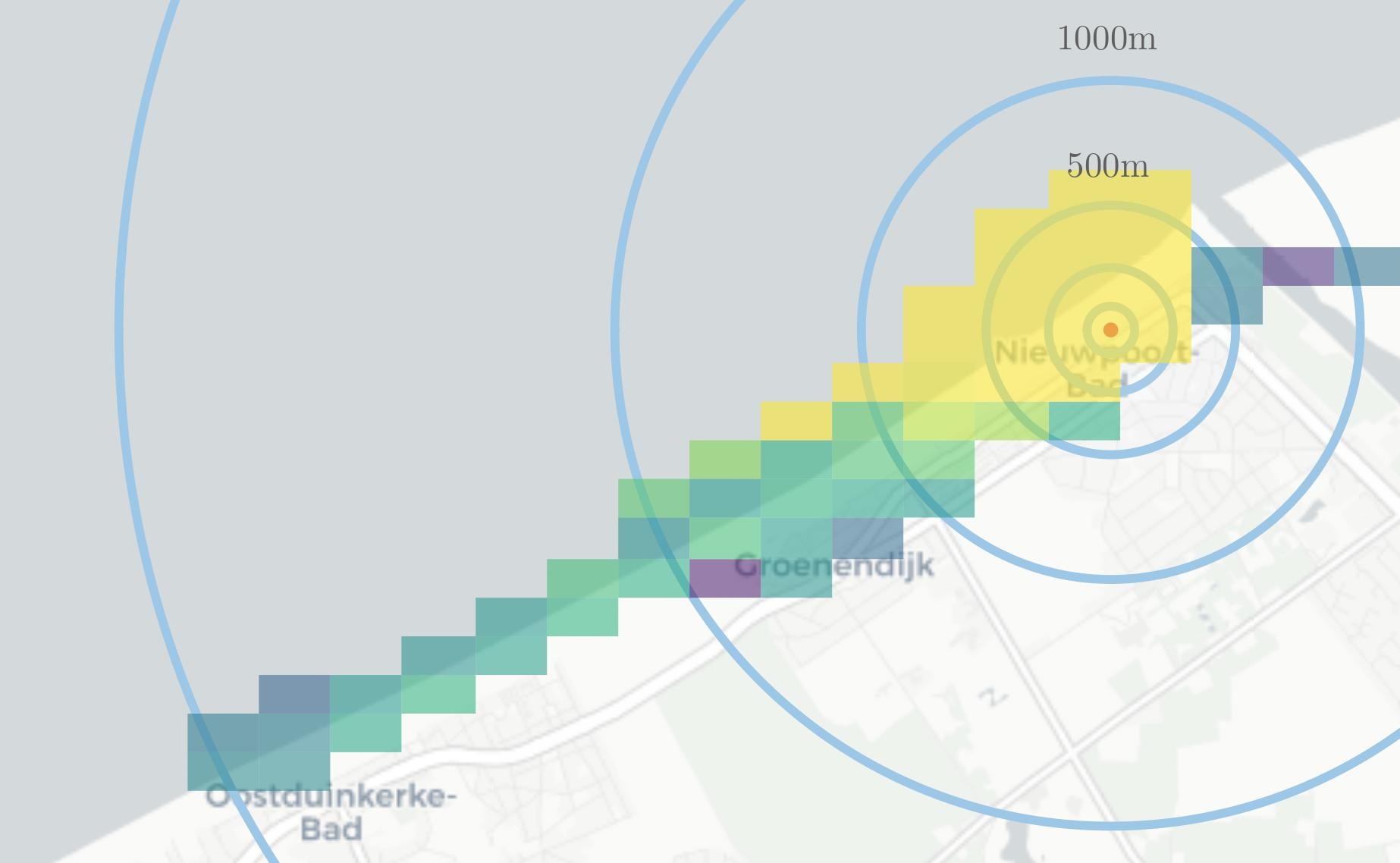}
        \caption{Coastal SNR}%
        \label{fig:heatmap-snr-coastal}
    \end{subfigure}
    \begin{subfigure}[b]{0.42\linewidth}
        \centering
        \includegraphics[width=0.9\linewidth]{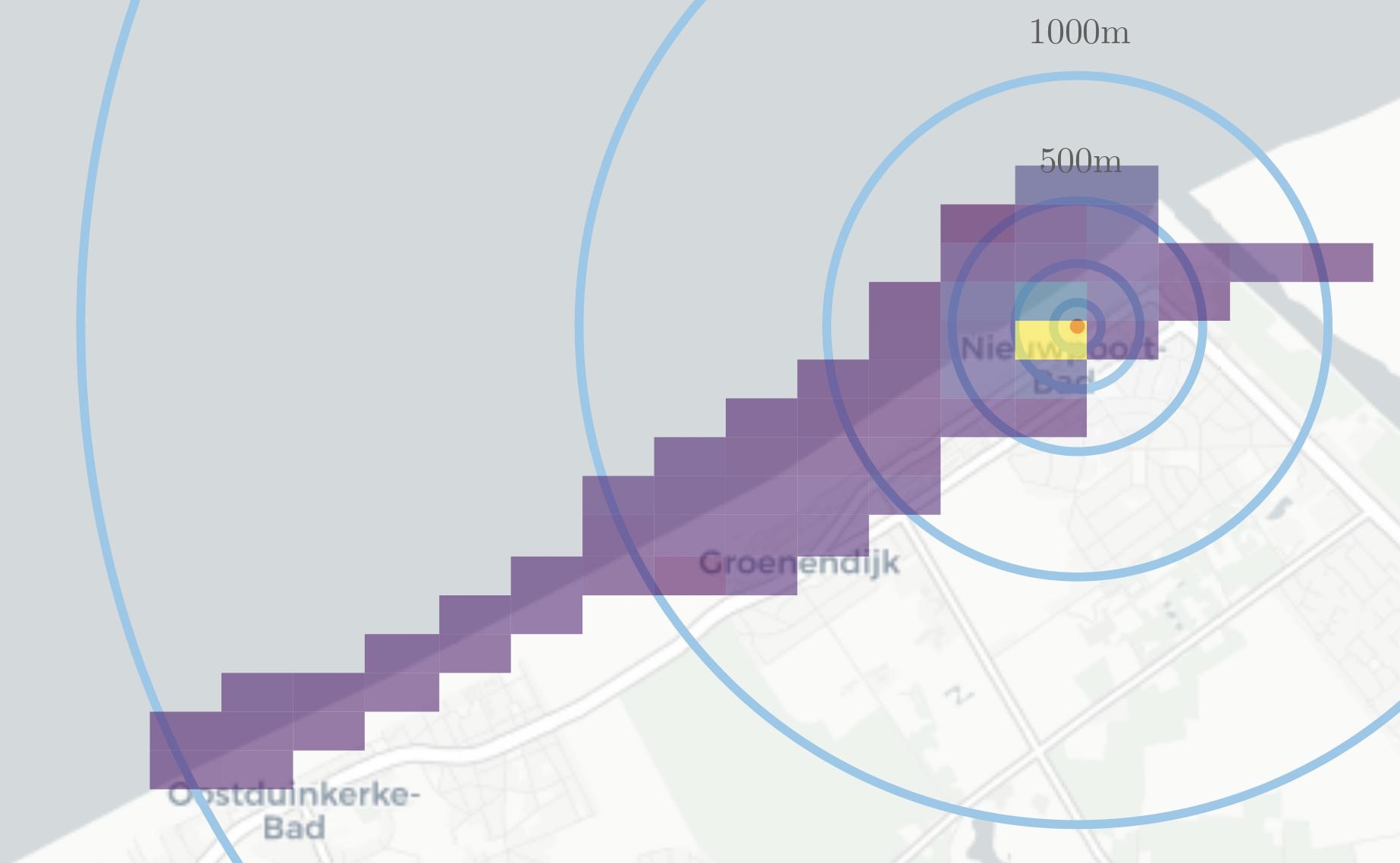}
        \caption{Coastal RSS}%
        \label{fig:heatmap-rss-coastal}
    \end{subfigure}

    \caption{Heatmap of signal-to noise-ratio and received signal strength for the coastal and forest environment.}\label{fig:heatmaps}
\end{figure}
\FloatBarrier

\section{Results of the Field Trials}\label{sec:results}
The \gls{rss} with respect to the Tx-Rx separation distance is illustrated in Figure~\ref*{fig:rss} for the urban environment, where we have gathered 10\,947 data points. The measurement campaigns conducted in the coastal and forested terrain resulted in 11\,017 and 4\,274 measurements, respectively.
A geographical representation of the \gls{rss} and \gls{snr} measurements is constructed by means of heatmaps and shown in Figure~\ref*{fig:heatmaps}. 

\textit{Quasi Fixed Average SNR due to the Spread Spectrum Technique: }
The heatmaps in Figure~\ref*{fig:heatmaps} demonstrate the impact of utilizing the adaptive spread spectrum technique. The \gls{snr} is boosted by increasing the \gls{sf}. 
In addition to improving the \gls{snr}, the sensitivity is increased yielding a higher link budget. 
However, transmitting with a higher \gls{sf} will not result in a higher \gls{rss}. Hence, the \gls{snr} does not decrease to the same degree as the \gls{rss} with increased distance.
Even in the urban scenario, a good average \gls{snr}, w.r.t the demodulation floor, is measured despite the low received signal strength. This validates the applicability of \gls{lora} \gls{p2p} in these scenario's. 

\textit{The Range Is Limited by the RSS: }
The experimental results indicate that the range is mainly limited by the \gls{rss} demodulation floor rather than the \gls{snr}. The transmit power, coding rate, spreading factor and other parameters will determine the maximum coverage. In our experiment, 
the coding rate and transmit power were constant and the noise is considered fixed. As a consequence, the largest coverage is obtained by employing the largest \gls{sf}, i.e. SF12.
The measurements show that the \gls{rss} sensitivity is reached before the \gls{snr} sensitivity for SF12. This confirms that the range is constrained by the \gls{rss}.

\textit{Long Range P2P links Are Possible with Low-Height Terminals: }
The coastal terrain clearly illustrates the favorable effect of a line-of-sight connection. The combination of line-of-sight and utilizing SF12 results in a coverage of over \SI{4}{\kilo\meter}; even with low-height terminals. 
Despite the unfavorable RF properties of the urban scenario and the densely forested terrain, a maximum range of \SI{1}{\kilo\meter} is still achieved.
\FloatBarrier


\section{Conclusions \& Future Work}\label{sec:conclusion}
In this paper we have reported on our study of \gls{lora} \gls{p2p} links to offer medium to long range IoT connectivity with easy and low cost communication infrastructure. 

We reported on the field trials in urban, coastal, and forest environments. The measurement data\footnote{\nolinkurl{github.com/DRAMCO/LoRaPointToPointMeasurementData}} as well as the developed hardware and software\footnote{\nolinkurl{github.com/DRAMCO/LoRaPointToPointModules}} are available open-source.
The resulting heatmaps demonstrate that a coverage up to \SI{1}{\kilo\meter} coverage is reached on obstructed urban links, while this extends to \SI{4}{\kilo\meter} in an open environment when transmitting at \SI{20}{\dBm} with transceivers at only \SI{1.5}{\meter} height.
This confirms \gls{lora} is a valuable technology for many IoT use cases. Especially in the context of farming, environmental monitoring, and logistics (e.g.\ in ports), where \gls{lpwan} coverage is often lacking, connectivity may hence be established with low investment and no subscription costs. 

Next, we will derive a path loss model for \gls{lora} \gls{p2p} connectivity with low-height terminals at both sides of the link. We anticipate a parametrizable model to represent different communication environments and to be used to assess other transmission schemes in the considered \SI{868}{\mega\hertz} band. This model could help both communication system designers and radio planning engineers in the future.

{\footnotesize%
\bibliography{bronnen.bib}{}%
\bibliographystyle{IEEEtranN}%
}
\end{document}